\documentclass[12pt]{iopart}
\pdfoutput=1
\usepackage{iopams}
\usepackage{graphicx}
\usepackage{hyperref}
\usepackage{color}

\begin{document}

\title{Effect of Noise on Patterns Formed by Growing Sandpiles}
\author{Tridib Sadhu$^{1}$\footnote{Present address : Department of Physics of Complex Systems, Weizmann
Institute of Science, Rehovot-76100, Israel.} and Deepak Dhar$^{2}$}
\address{ Department of Theoretical Physics,\\
Tata Institute of Fundamental Research, Mumbai-400005, India.}

\eads{\mailto{$^{1}$tsadhu@gmail.com}, \mailto{$^{2}$ddhar@theory.tifr.res.in}}

\date{\today}

\begin{abstract}
We consider patterns generated by adding large number of sand grains
at a single site in an abelian sandpile model with a periodic initial
configuration, and relaxing. The patterns show proportionate growth.
We study the robustness of these patterns against
different types of noise, \textit{viz.}, randomness in the point of 
addition, disorder in the initial periodic configuration, and
disorder in the connectivity of the underlying lattice. We find that
the patterns show  a varying degree of robustness to addition of a small
amount of noise in each case. However, introducing stochasticity in
the toppling rules seems to destroy the  asymptotic patterns completely, even for a weak
noise. We also discuss a variational formulation of the pattern selection problem in growing
abelian sandpiles.
\end{abstract}

\submitto{ Journal of Statistical Mechanics: Theory and Experiment}
\maketitle

\section{Introduction}
The problem of how a large animal develops from a single cell has been
a central problem in  biology. A somewhat simpler,  but nontrivial,
problem  is how  a baby animal grows into an adult, increasing the
total body mass by two or three orders of magnitude, while different
parts of the body keep roughly same proportions during growth. We
will refer to this property as \textit{proportionate growth}. Our work
has been motivated by trying to construct minimal physical models with
this property.

A simple example of proportionate growth in a non-biological context
is a dew drop on a windowpane. Its shape may be approximately
described as a spherical frustum, where the contact angle with the
glass surface is determined by the surface tension. As it takes water
from the super-saturated air in the surrounding, it grows in size, and
shows nearly proportionate growth. However, it is not easy to
construct models showing proportionate growth in patterns with
substructures. In fact, no other model of growth studied in
physics literature so far, shows
this property. In the well studied Eden model \cite{eden},
diffusion-limited aggregation \cite{dla}, invasion percolation
\cite{invasion}, or the Kardar-Parisi-Zhang type models \cite{kpz},
growth occurs in the outer ``active regions'', whereas the inner core once
formed, remains essentially frozen afterwards.

It seems reasonable that a proportionate growth would require some
central regulation or a long-distance communication and coordination
between different parts of the structure. For an animal growth this is
certainly true. The growth is orchestrated by the turning on and off
of different regulatory enzymes and chemicals, ultimately determined
by the genetic program  encoded in the cell's DNA. It is interesting
that, such growths can be achieved in a model system with components of
much lower complexity, \textit{i.e.}, a cellular automaton model with only a few
states per site.  In an earlier work \cite{epl}, we have studied the Abelian
Sandpile Model (ASM) as
the prototypical model of proportionate growth. The
patterns are formed by adding large number of sand grains at a single
site on a periodic initial configuration (also referred to as
``background''), and letting the system relax to a stable configuration. We were able to characterize 
the full pattern analytically in one simple case. The effect of adding absorbing
sites, or multiple centers of growth was studied in \cite{jsp}.

Real biological growth occurs in a fairly noisy environment. While
deterministic cellular automaton models with simple toppling rules
cannot be considered realistic  biologically, it is still interesting
to ask whether the patterns produced by growing sandpiles are robust
against introduction of a small amount of noise. In the presence of
noise, an analytical study of this problem is quite difficult.  The
techniques used in \cite{epl} to characterize the pattern exactly, no
longer work, as they depend on the potential function in each patch
being a quadratic function of the coordinates (see \sref{sec:prel}).
The work reported here is exploratory, and mainly numerical. However, the
fact that the patterns show some degree of robustness, even in the
presence of noise, strongly suggests that this is not a special
property related to the exact solubility of the ASM, and a more general
macroscopic description of pattern formation and pattern selection in
this problem, not requiring an exact solution, should be
possible. We discuss how the ``least action principle'' for
ASMs could provide a possible framework for understanding pattern
formation in our problem, as the variational formulation provides
a quantitative criterion for pattern selection.
\begin{figure}
\begin{center}
\includegraphics[width=6.0cm]{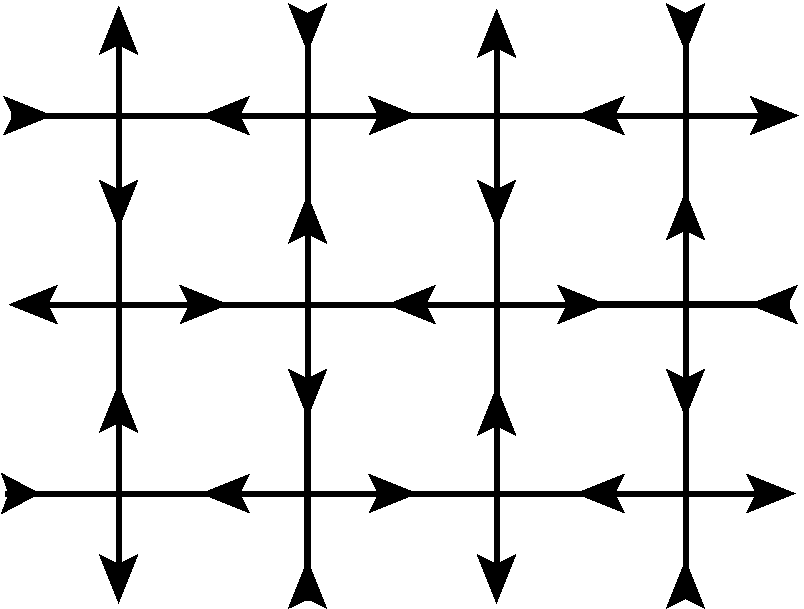}
\caption{The F-lattice: A square lattice with directed edges whose
directions are assigned periodically as shown.}
\label{fig:flattice}
\end{center}
\end{figure}

We have studied the robustness of these patterns against different
types of noises, namely, random fluctuations in the position of the point of
addition of grains, disorder in the periodic background
configuration of heights, and  disorder in the connectivity of the
underlying background lattice. We find that the patterns show a varying
degree of robustness to addition of a small amount of noise in each
case. However, introducing stochasticity in the toppling rules seems
to destroy the asymptotic pattern completely, even for a weak noise.

The spatial patterns formed in sandpile models were first studied by
Liu \textit{et.al.} \cite{liu}. The asymptotic shapes of the boundaries of
sandpile patterns produced by adding grains at single site on
different periodic backgrounds was discussed in a later work by Dhar \cite{dhar99}. Borgne
\textit{et.al.} \cite{borgne} obtained some bounds on the
rate of growth of these boundaries, and later these bounds were
improved by Fey \textit{et.al.} \cite{redig} and Levine
\textit{et.al.} \cite{lionel}. The first detailed analysis of
different periodic structures found in the sandpile patterns was discussed by
Ostojic \cite{ostojic}. An extensive collection of centrally
seeded sandpile patterns on different lattices, with high resolution
images, can be seen in
\cite{wilson}. Other special configurations in the ASMs, like the identity \cite{borgne,identity,caracciolo} or the
stable state produced from special unstable states \cite{liu}, also show complex
structures , which share common features with the single
source patterns studied here.

This paper is organized as follows. In \sref{sec:prel}, we define the
models precisely, and introduce the scaled excess
density function and the scaled toppling function. These functions give a quantitative
characterization of the patterns. In \sref{sec:addition}, we discuss
the robustness of the patterns against small fluctuations in the
position of the point of addition. In \sref{sec:bkg}, we discuss the
effect of noise in the background configuration. In
\sref{sec:quenched}, we discuss the effect of disorder in the underlying lattice on
which the growth occurs. This is modeled by a quenched disorder in
the toppling rules. We find that, the patterns are quite sensitive to
changes in the toppling matrix. In \sref{sec:manna}, we discuss the
effect of noise due to fluctuations in the toppling process,
\textit{i.e.}, at each
toppling, there is  a small probability that the grain transfer occurs
in a direction not given by the toppling rule. We find that, even a
very small amount of noise in the toppling rules completely wipes out
the asymptotic pattern. Finally, in the concluding \sref{sec:discussion}, we 
discuss  the `least action principle' for the ASM, and suggest that
it could provide a basic framework for understanding pattern formation in these systems.
\begin{figure}
\begin{center}
\includegraphics[width=7.5cm]{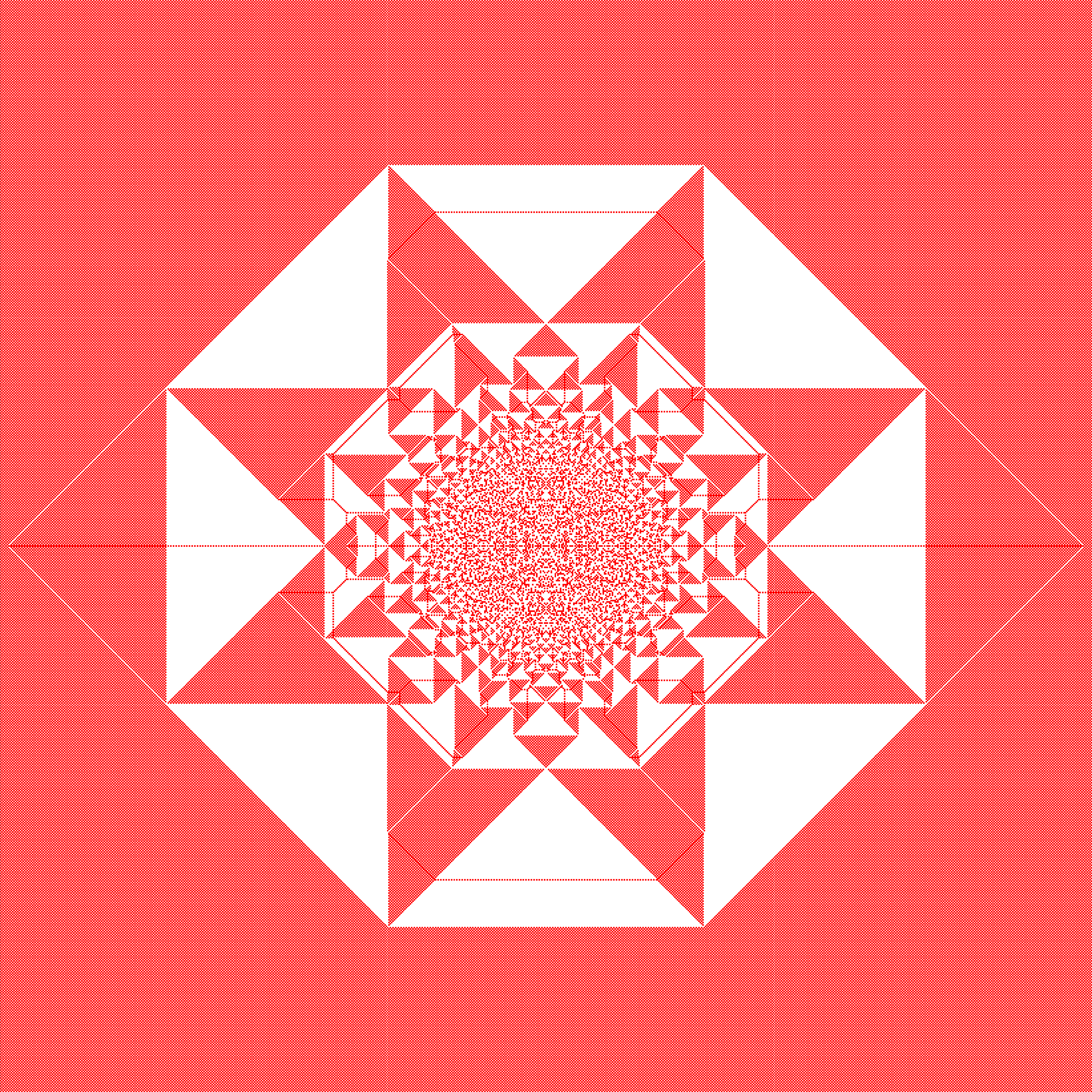}
\includegraphics[width=7.5cm]{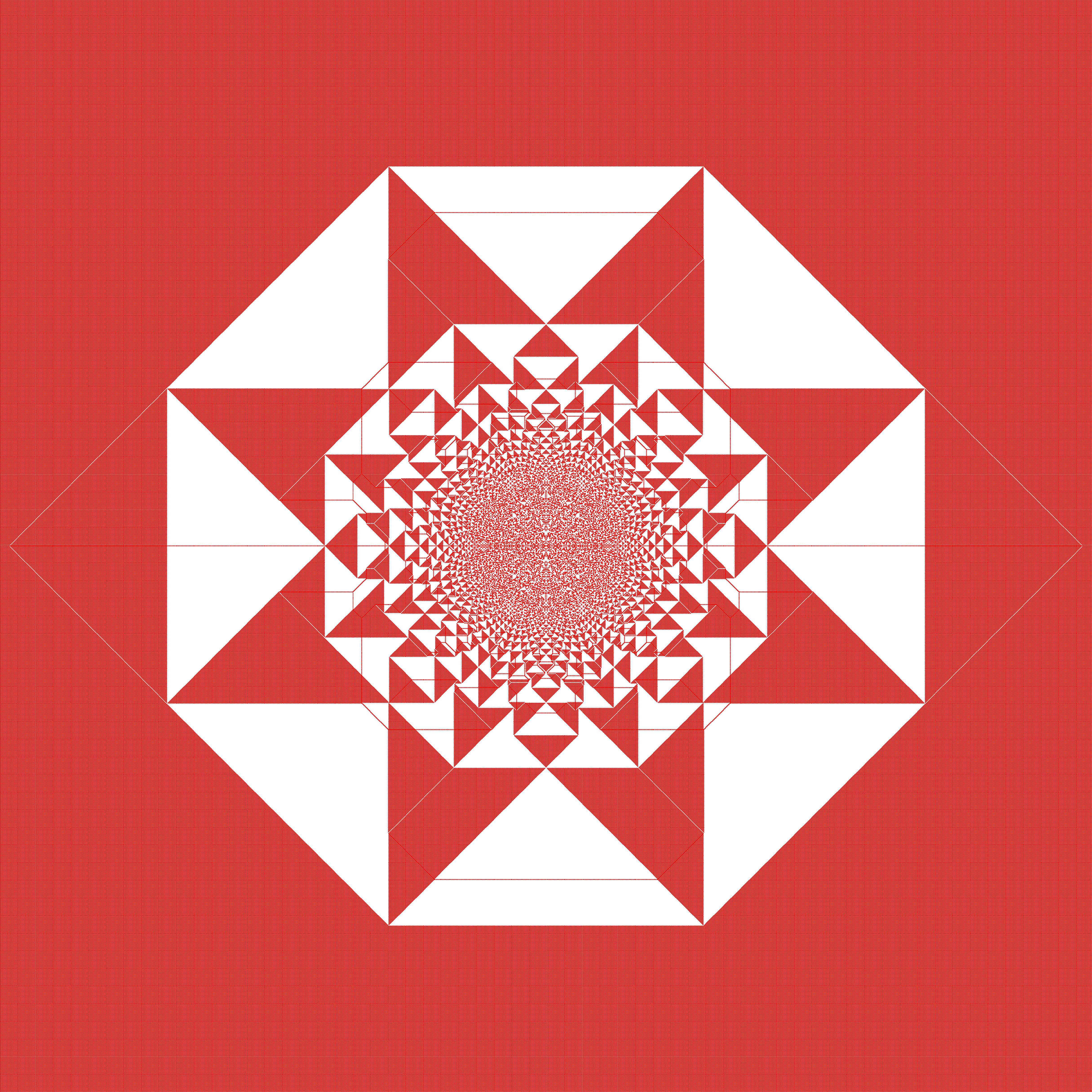}
\caption{The patterns formed in the ASM defined on the F-lattice with
checkerboard background of heights $0$ and $1$. These two patterns
correspond to $N=80,000$ and $N=320,000$ grains, respectively. Color code:
Red$=0$, White$=1$. For comparison, the size of the first pattern has been enlarged by a factor
two.}
\label{fig:prop}
\end{center}
\end{figure}

\section{Preliminaries and definitions}\label{sec:prel}
For our numerical studies, we have used two model systems (Here the
term ``model system'' is used as in biology literature: the fruit fly
is a model animal, and biological functions in other animals are
qualitatively similar).

The first model is defined on an infinite square lattice with directed
edges, such that at each site there are two edges coming in, and two
going out (\fref{fig:flattice}). This directed square lattice is
called the F-lattice. At each site $\mathbf{x}$, there is a non-negative integer
variable $z\left(\mathbf{x}\right)$, called the number of sand grains 
at $\mathbf{x}$,  also called the height of the sandpile
at that site. Any site with height
greater than $1$ is said to be unstable, and it topples by
transferring exactly two grains in the direction of outward arrows
from that site. We start with an initial configuration in which the
heights $0$ and $1$ form a checkerboard pattern. At each time step, we
add a grain at the origin, and let the resulting configuration relax
until all sites are stable. After $N$ grains have been added, with $N$
large, we find that the heights form an intricate and beautiful pattern,
whose size grows as $N$ increases. The resulting patterns for $N=
80,000$ and $320,000$ are shown in \fref{fig:prop}. Note that, what
appears to be solid red region in the figure due to low resolution, is
actually a checkerboard pattern of alternate red and white squares.
Details may be seen by zooming in. We see that the two scaled patterns
are the same, except that the smaller patches close to the center of the pattern are
resolved better in the second.

\begin{figure}
\begin{center}
\includegraphics[width=8.0cm]{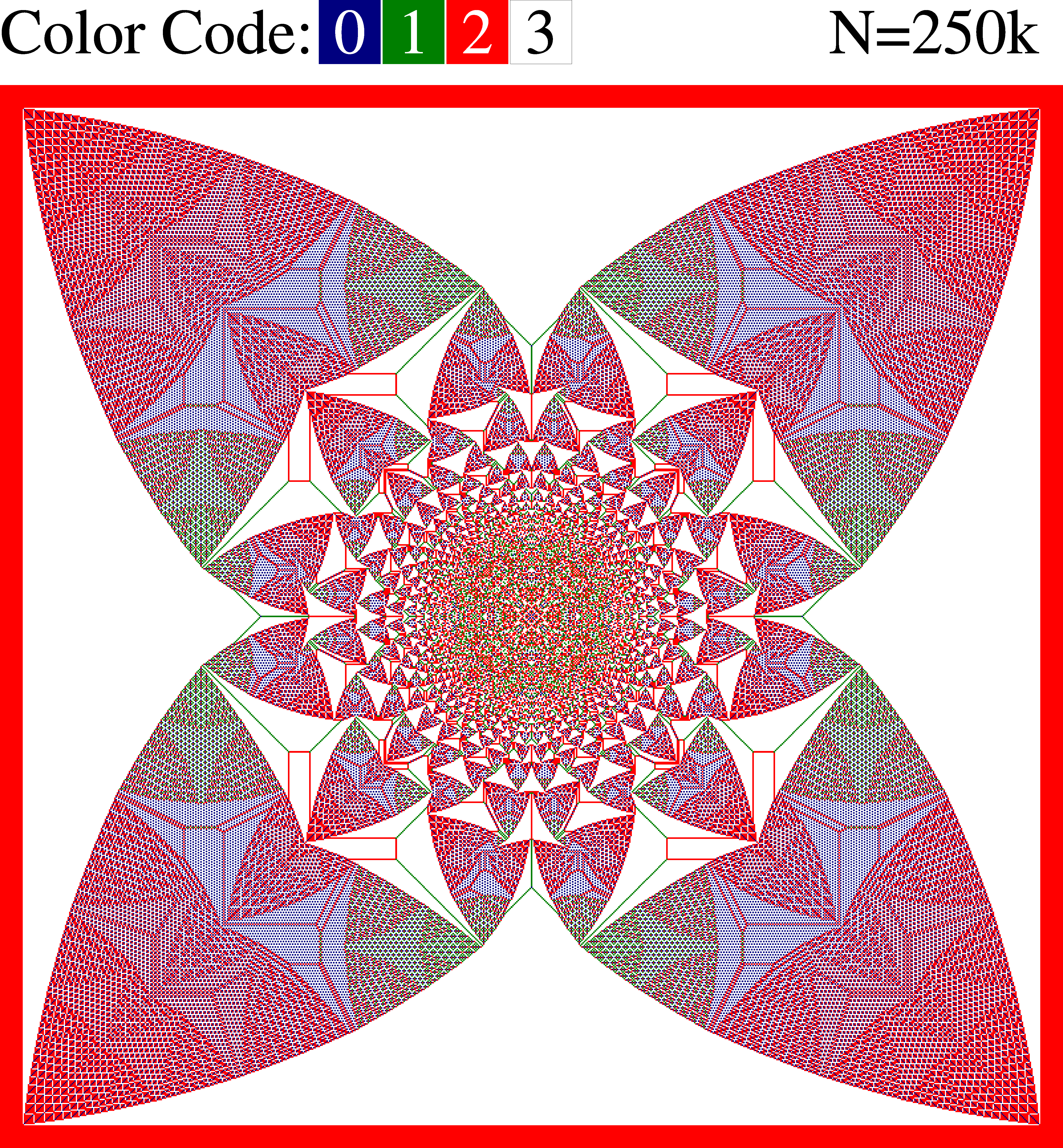}
\caption{A pattern produced on a square lattice with a background of
height $2$ at every site.}
\label{fig:btw}
\end{center}
\end{figure}

The second model system is the ASM on an undirected infinite square
lattice. We define the ASM on this lattice as follows: in a stable
configuration, all sites have heights less than $4$. Any site where the height is
greater than $3$ is said to be unstable, and it relaxes by transferring $4$ sand
grains from that site, one to each of the four nearest neighbors. We
start with an initial configuration where the height at each site is
$2$. The stable configuration after adding $N= 250,000$ grains at the origin is
shown in \fref{fig:btw}.

One can consider patterns obtained when the initial configuration has a
different periodic structure. For the undirected square lattice, when
the initial configuration is a periodic arrangement of heights, in which
each site has height less than or equal to $2$, one finds a pattern in
which the diameter of the pattern grows as $\sqrt{N}$ \cite{denboer}.
For the F-lattice, there are some backgrounds on which the growth
of the pattern is faster than $\sqrt{N}$ \cite{triangular}. In all the cases studied
so far, if there are no infinite avalanches, the patterns show
proportionate growth. Although we do not have a rigorous proof of this
important property, there is good numerical evidence, and we shall assume it in the following.

A key observation is that, for large $N$, the patterns in
\fref{fig:prop} or \fref{fig:btw} may be seen as a union of disjoint
patches, each of which occupies a non-zero fraction of the area of the full
pattern, and the arrangement of heights inside a single patch is exactly
periodic. We denote the diameter of the pattern by $\Lambda$, which
may be defined as the width of the smallest square enclosing the
pattern.
We define reduced coordinates $\xi =x/\Lambda$ and $\eta = y/\Lambda$. The
local excess density of grains  $\Delta \rho( \xi,\eta)$ is  defined
as the difference in the density of grains in the final and initial
patterns, in a small neighborhood of the point $(\xi,\eta)$ in the reduced
coordinates. We specify the asymptotic pattern in the
limit of large $N$, by specifying the function $\Delta \rho(
\xi,\eta)$. From the fact that inside each patch, there is a periodic
pattern of integer heights, it follows that the excess density $\Delta
\rho(\xi,\eta)$ is a rational constant for each patch.

It is useful to define a function $\Phi(\xi,\eta)$ as the
scaled number of topplings at the  site with reduced coordinates $(\xi,
\eta)$. Let $T_N(x,y)$ be the number of topplings at site $(x,y)$, after
adding $N$ grains and relaxing the system completely. We define the function
\begin{equation}
\Phi(\xi,\eta) = \lim_{N \rightarrow \infty} \frac{1}{\Lambda^2}T_N([\xi \Lambda],[y \Lambda]),
\end{equation}
where $[x]$ denotes the integer nearest to $x$.  The conservation of
number of grains implies that the potential function satisfies the Laplace's
equation \cite{epl}
\begin{equation}
\nabla^2 \Phi(\xi,\eta) = -\delta(\xi,\eta) + \Delta \rho(\xi,\eta).
\end{equation}

\begin{figure}
\begin{center}
\includegraphics[width=7.5cm]{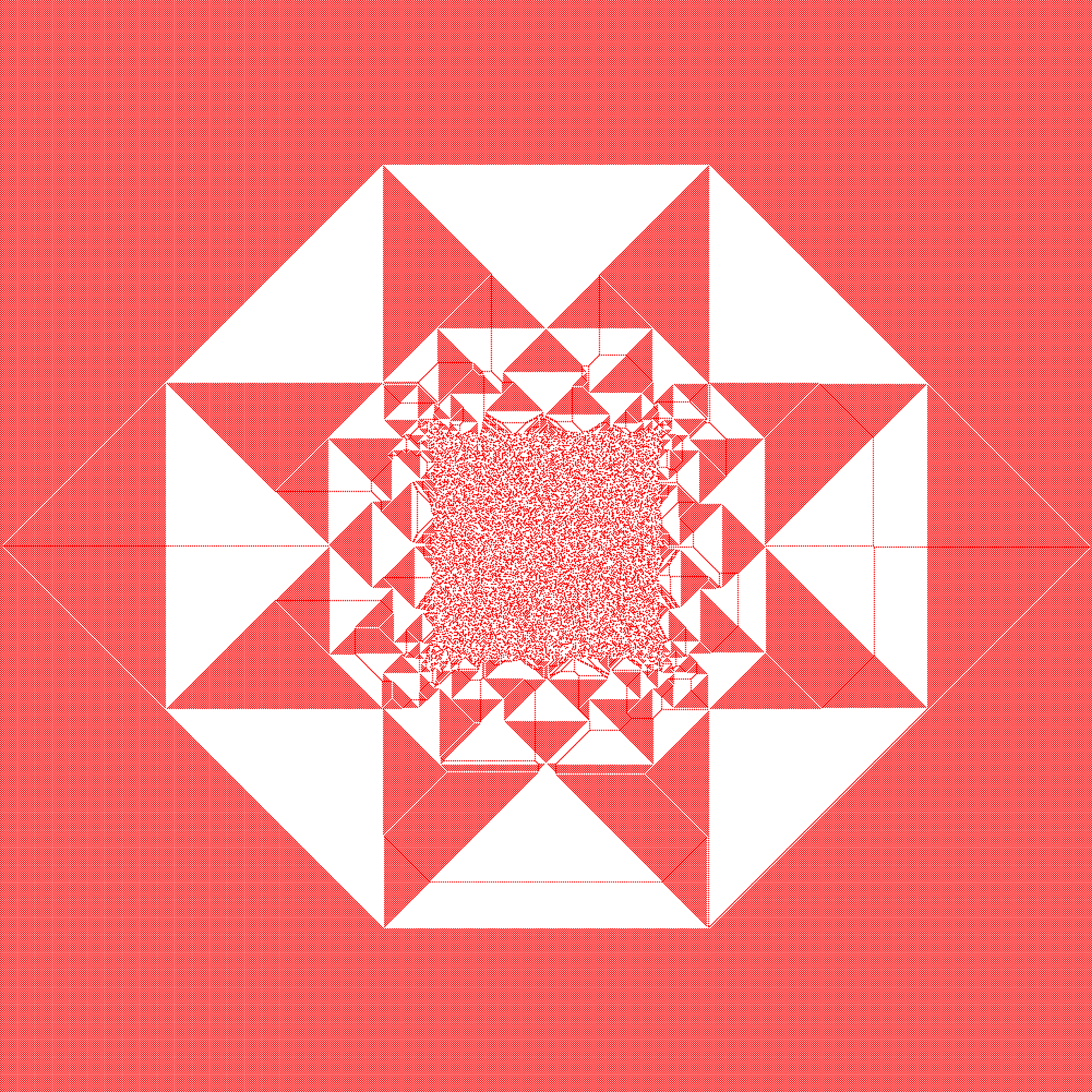}
\includegraphics[width=7.5cm]{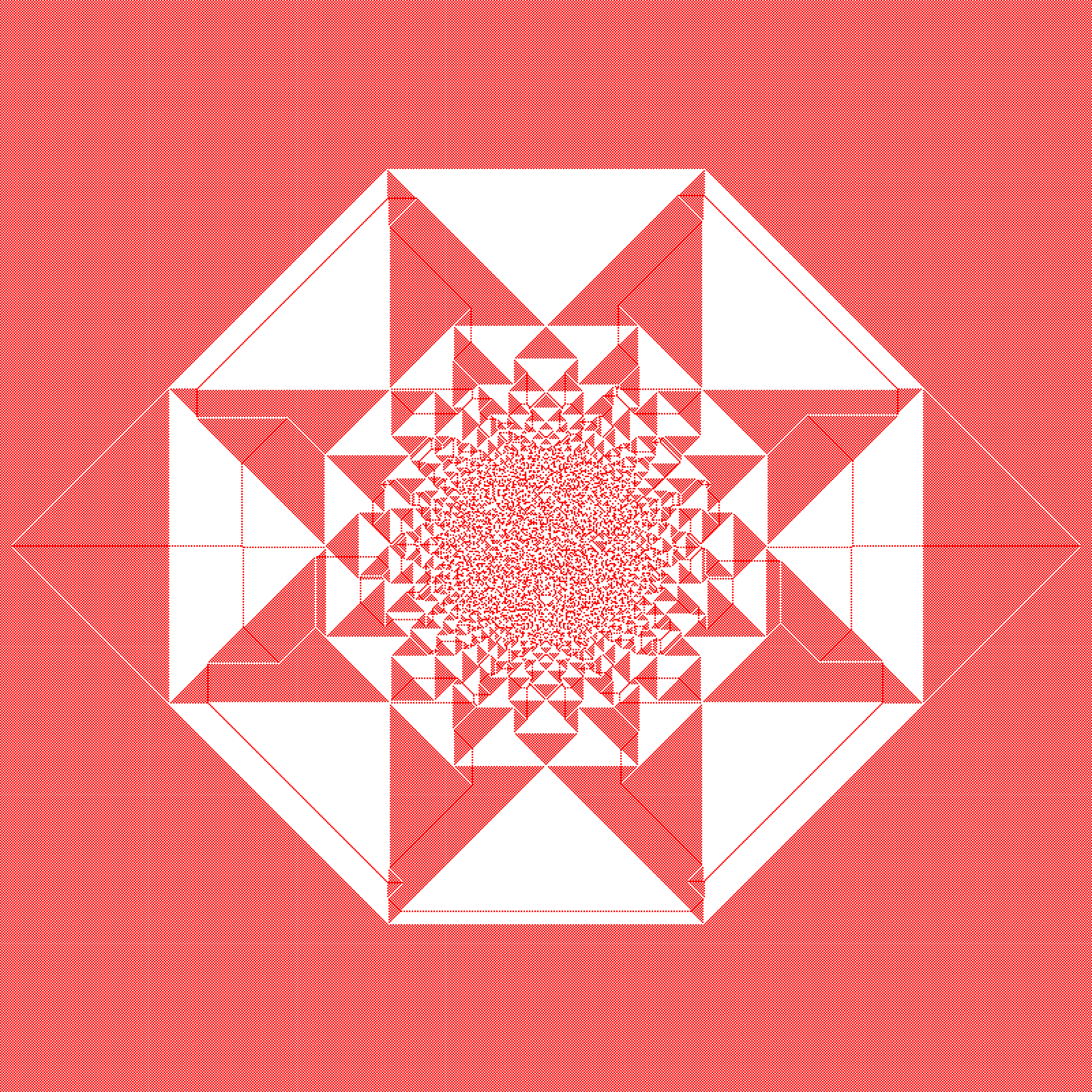}
\caption{The patterns produced on the F-lattice by adding sand grains at
sites randomly chosen from a square region of area equal to $9\%$ and
$1\%$, respectively, of that of the corresponding final patterns. The
initial configuration is with checkerboard distribution of
heights $0$, and $1$. Color code: Red=0, white=1.}
\label{fig:30perc}
\end{center}
\end{figure}
In an electrostatic analogy, we can think of  $\Phi(\xi,\eta)$ as
the potential produced by a unit positive point charge at the origin,
and an areal charge density $-\Delta \rho(\xi,\eta)$. We shall refer to
$\Phi(\xi,\eta)$ as the potential function hereafter. In each periodic patch,
the potential $\Phi(\xi,\eta)$ is a quadratic function of the
coordinates $\xi$ and $\eta$ \cite{ostojic}. For the pattern on the F-lattice, it was
shown that the coefficients of the quadratic terms are simple
rational numbers, while the linear terms can be determined by the
condition that the potential and its derivative are continuous
functions at the boundaries where two patches meet. This allowed us to
characterize the asymptotic pattern completely \cite{epl}.

\section{Effect of fluctuations in the site of addition}\label{sec:addition}
The patterns studied so far are produced by adding one grain at each time step,  
at a fixed site (the origin). Now, we
consider how these change when the site of addition fluctuates in time at
random. To be more specific, we add $N$ grains by randomly
distributing them among sites within a small square centered at the
origin. The size of the square is  chosen to be of length $\epsilon
\Lambda$, with $\epsilon<1$. The background is a checkerboard
distribution of heights $1$ and $0$. 
\begin{figure}
\begin{center}
\includegraphics[width=8.0cm]{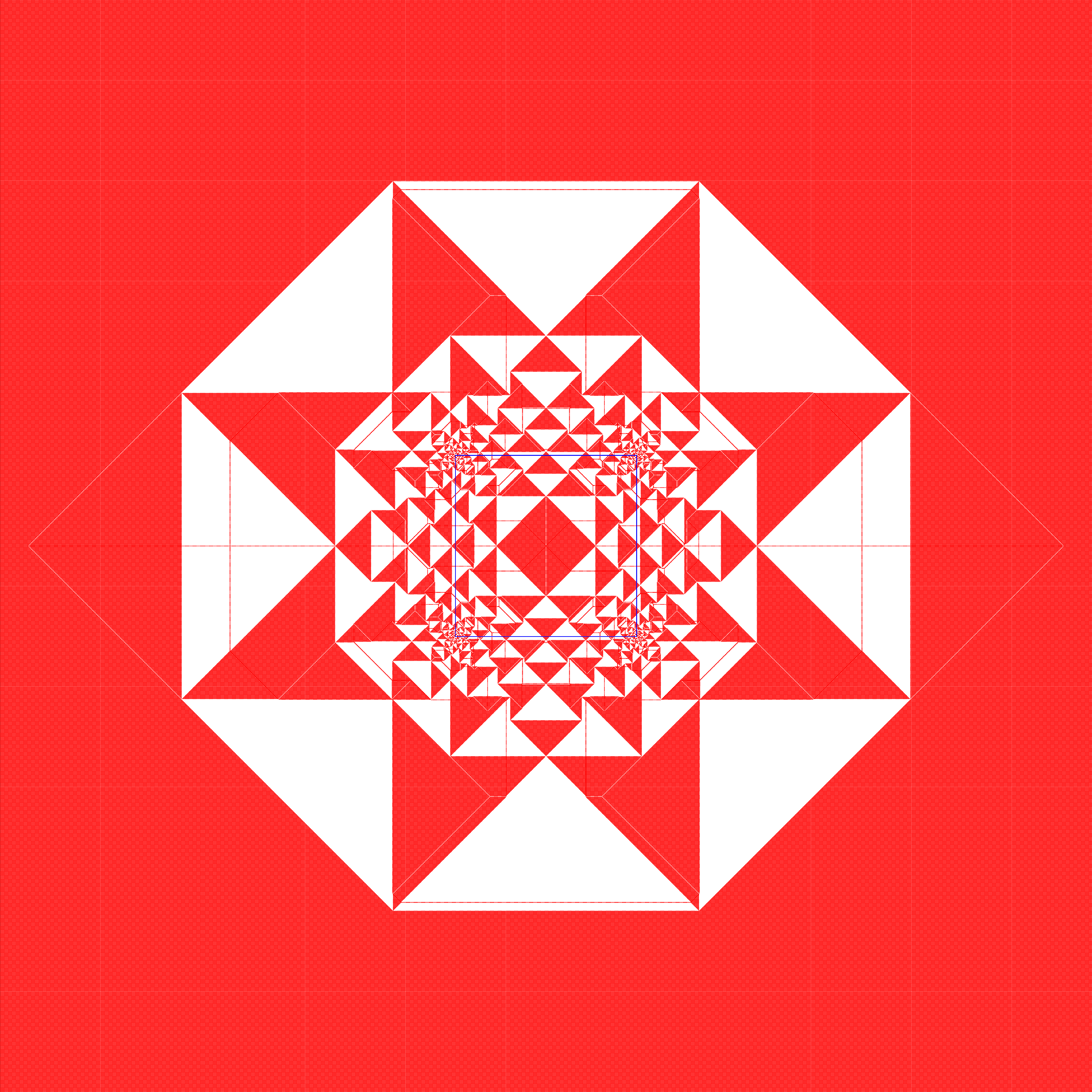}
\caption{A pattern similar to the one in \fref{fig:30perc}, but this
time the grains are added uniformly, four at every site inside the
square of width $160$ lattice units. The diameter of the full pattern
is $592$. Color code: Red=0, White=1.}
\label{fig:uh}
\end{center}
\end{figure}

We have shown the pattern for $\epsilon=0.3$ and $N=120,000$ in
figure \ref{fig:30perc}a, and
the pattern corresponding to $\epsilon=0.1$ and $N=75,000$ in figure
\ref{fig:30perc}b. We see that the patches away from the boundary of
the square region of addition, are identical to that of the
single source pattern (compare with \fref{fig:prop}). That is, the patches at
the outer layers in the pattern, are not much affected by this change. Only near the
center, within a distance of order $\epsilon$, we see a change.
Near the center, the dense set of patches of decreasing size
is smeared out, and there are no periodic quadratic patches left.
However, there are new accumulation points of patches, which develop
at the four corners of the square region of addition.
As more grains are added, finer patches appear in a way that their number
would become infinite in the asymptotic limit of large $N$. For small
$\epsilon$, the shape of the outer parts of the pattern shows only a
weak dependence on its value.

The pattern changes in an interesting way, when the addition of
grains are not random. The pattern produced by uniformly adding
$m=4$ grains at every site inside a square of size $2l\times 2l$ is
shown in figure \ref{fig:uh}. The boundary of the square is indicated
by a solid blue line. The pattern outside and away from the square, looks very
similar to the pattern corresponding to random addition. The most significant change is
inside the square, where there are well defined periodic patches.
Again, there is no accumulation point of tinier and
tinier patches at the origin. As we increase $l$, keeping $m$ fixed, more and more patches appear near the corner of
the square. It seems that their number would tend to infinity for
large $N$.
\section{Randomness in the initial configuration}\label{sec:bkg}
\begin{figure}
\begin{center}
\includegraphics[width=7.5cm]{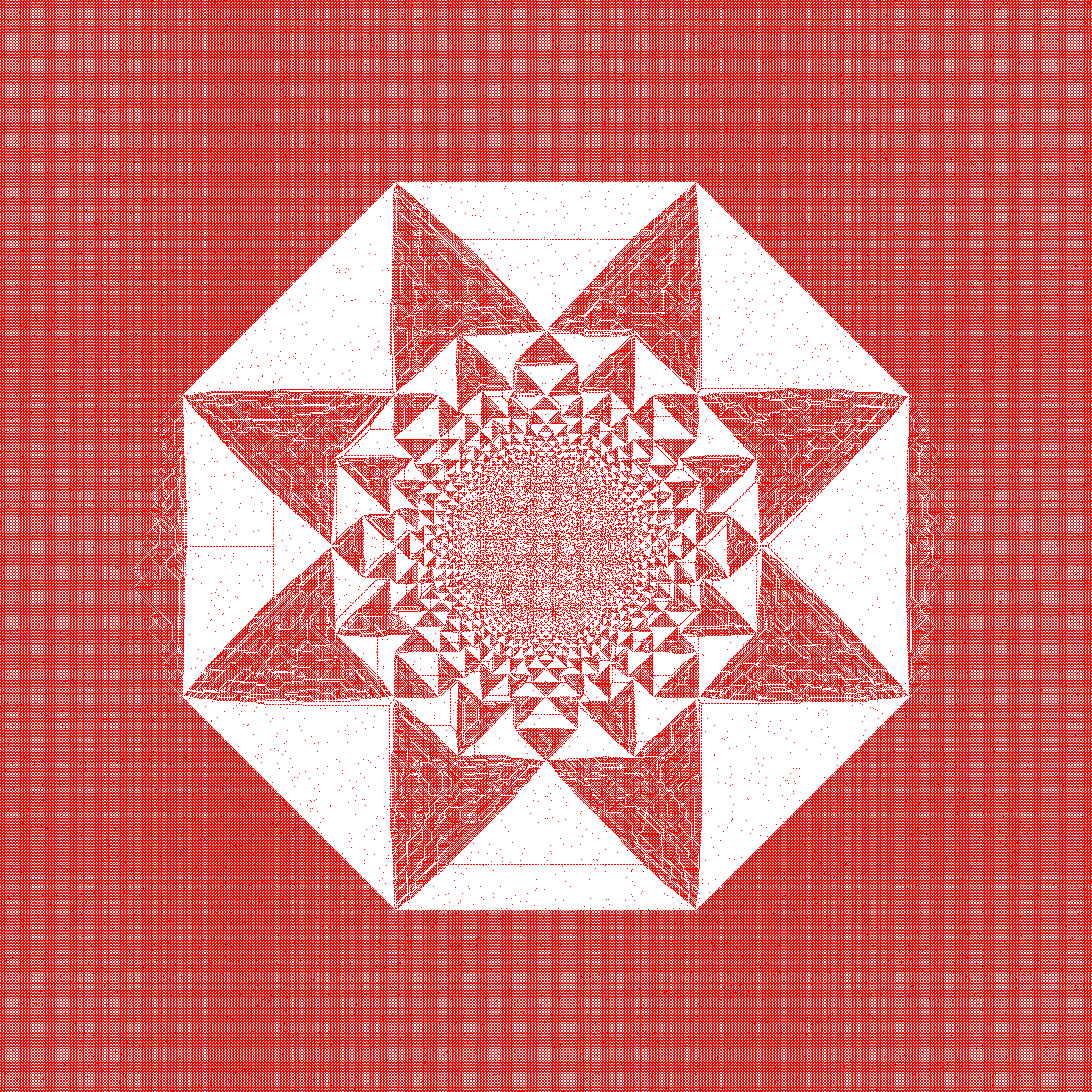}
\includegraphics[width=7.5cm]{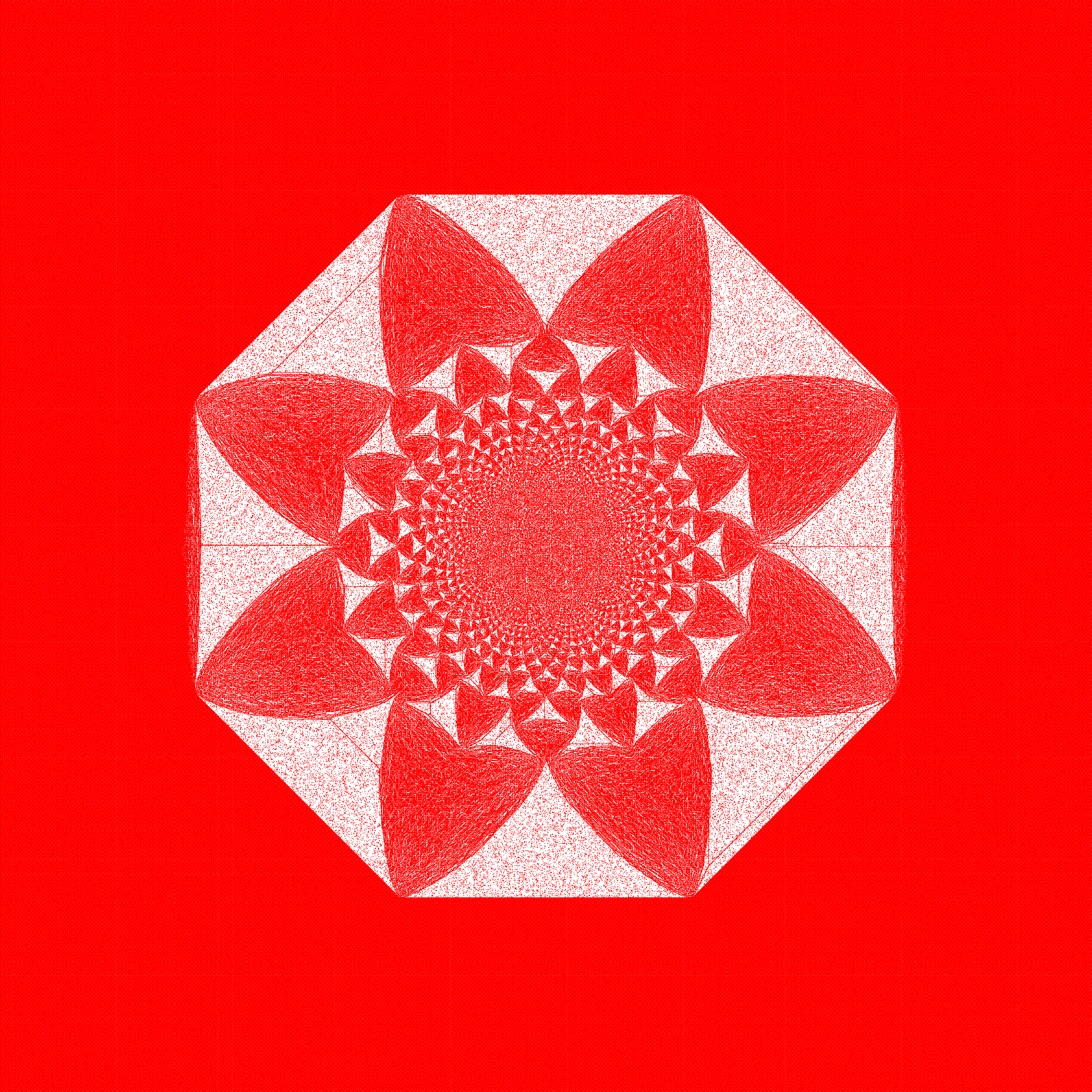}
\caption{The patterns produced on the F-lattice by adding $N=228,000$ and
$N=896,000$ grains at a single site on a background of mostly checkerboard
distribution, except height $1$ at $1\%$ and $10\%$ of the sites,
respectively, are replaced by height zero. Color code: Red=0, White=1.}
\label{fig:idtype1}
\end{center}
\end{figure}
The patterns show a significant amount of robustness to the noise in
the background. The least effect of change in the background on
F-lattice occurs if some heights $1$ are replaced by heights zero.

This is easy to see using the abelian property of the
ASM. Let $C$ be the initial height configuration and $D$ be the final configuration produced
by adding $N$ grains at the origin. Consider a particular site $i$, which  has
height $1$ in both $C$ and $D$. Let the configurations obtained from
$C$ and $D$ by changing the height at this site from $1$ to $0$ be
called $C'$ and $D'$, respectively. Then, if $C'$ and $D'$ contain no forbidden sub-configurations
\cite{dharphysica}, using the abelian property one can show that addition of $N$ grains in $C'$ would give relaxed configuration $D'$. Also the toppling function is same in both the
cases.  

Thus we expect that a very  small concentration of $1$'s replaced by
$0$'s will have only a small effect. 
This expectation is verified numerically. The
patterns on the F-lattice corresponding to backgrounds with $1\%$ and $10\%$ noise in
the background are shown in \fref{fig:idtype1}a and
\fref{fig:idtype1}b, respectively. We see that the qualitative
structure and placement of different patches is unaffected in the
low noise case. In particular, there are only two types of patches, and the relative positions and
sizes of the larger patches are not changed much. The excess density is uniform within
each patch, and jumps sharply across clearly defined patch
boundaries. The outer boundary of the pattern, separating the
red region outside and the eight largest white patches, seems to remain a nearly
perfect octagon, but with slightly rounded off corners. However, other
patch boundaries are no longer straight lines in the presence of noise, and
a significant curvature in the patch boundaries is clearly seen in
patterns for larger noise.
\begin{figure}
\begin{center}
\includegraphics[width=7.0cm]{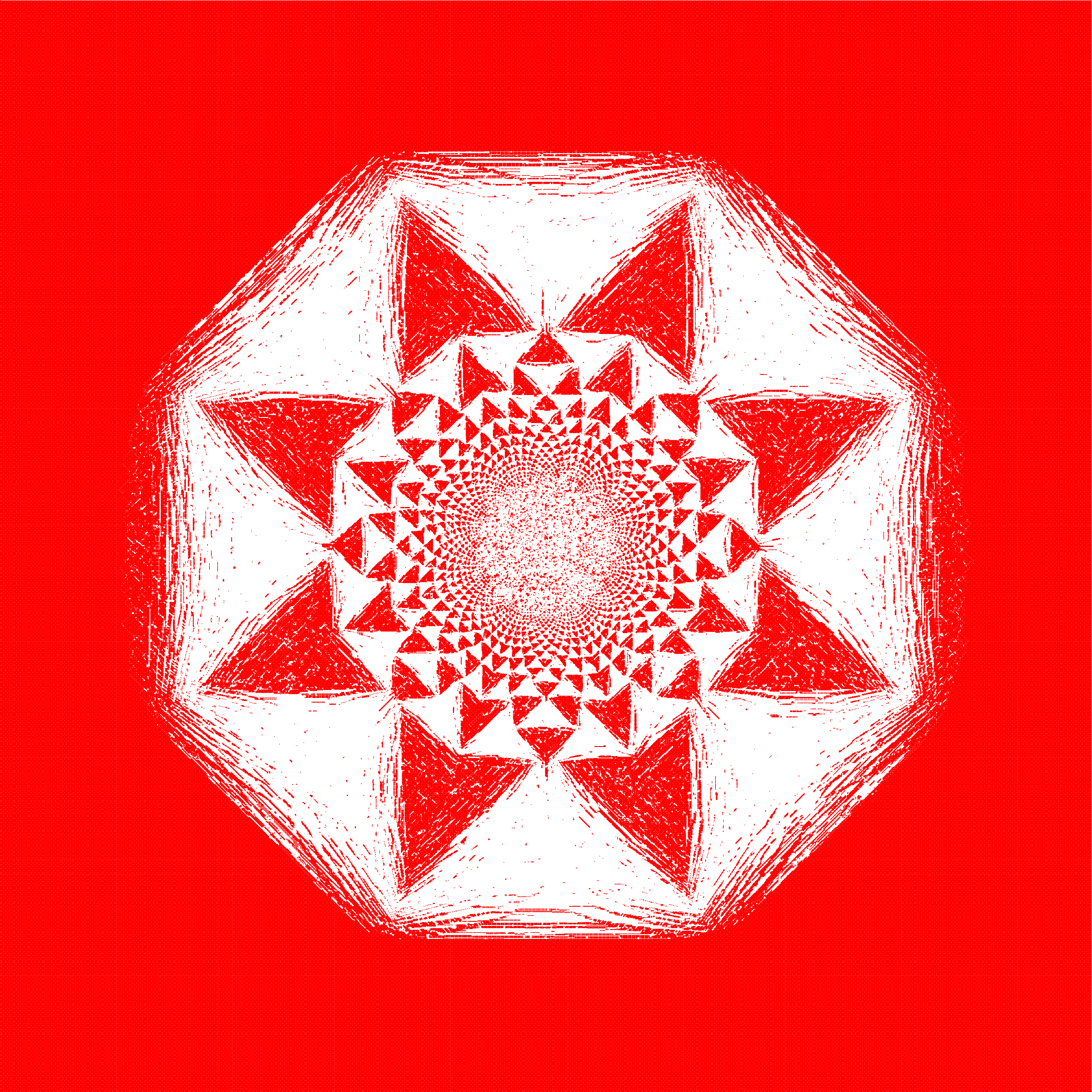}
\includegraphics[width=7.0cm]{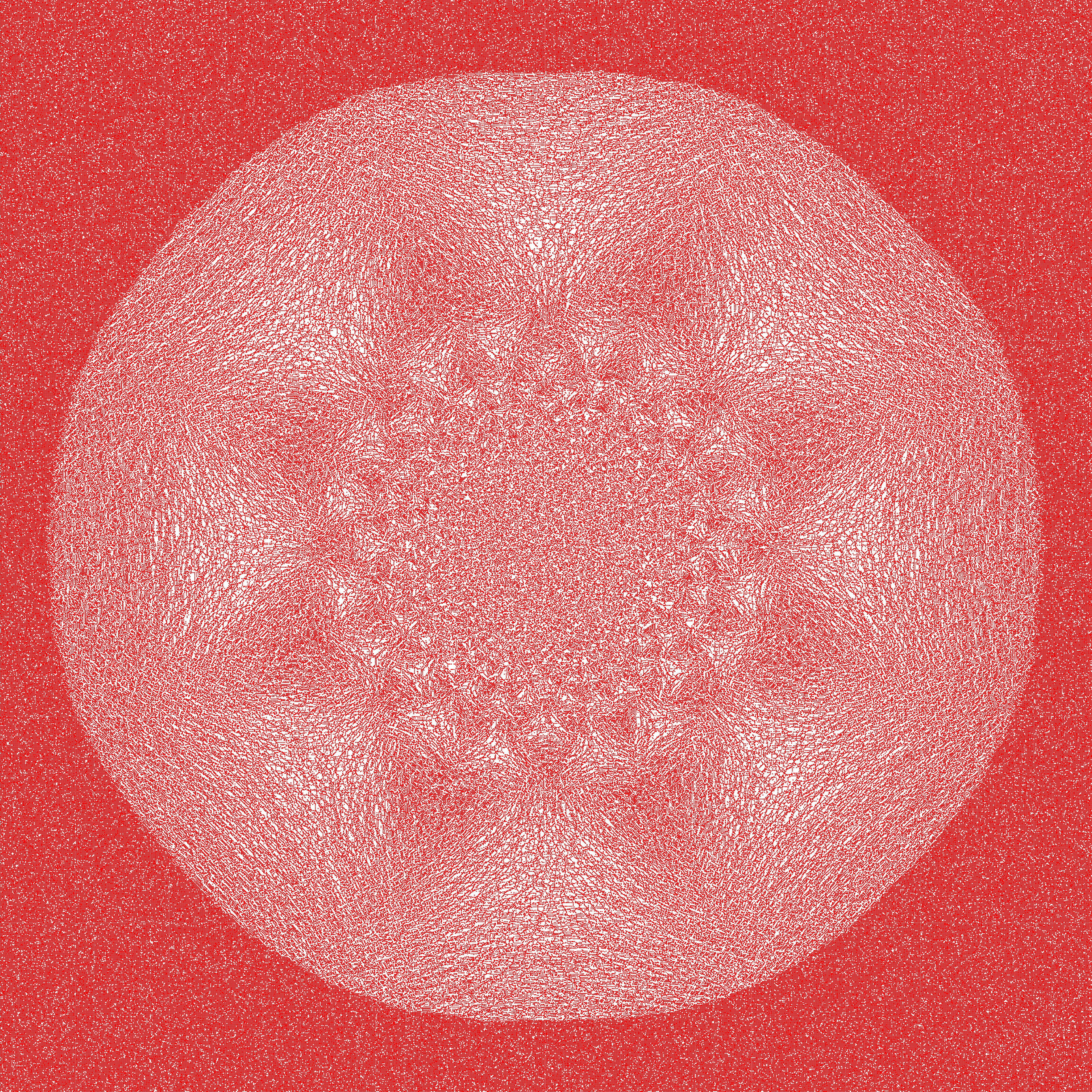}
\caption{The patterns produced on a mostly checkerboard background,
except heights at $1\%$ and $10\%$ sites, respectively, are flipped.
Color code: Red=0, White=1.}
\label{fig:ID1}
\end{center}
\end{figure}

Even for the relatively large noise value (\Fref{fig:idtype1}b), the basic 
eight petal structure  of the pattern without noise is clearly visible. 
However with increase of defect density, the relative area of the
dense patches (white color) decreases. Also the corners of the outer boundary of the
pattern becomes smoother, and for defect density close to $50\%$, the
pattern becomes a circle with a single aperiodic patch inside.

The patterns are more sensitive to the changes in heights $0$'s to $1$'s.  In
figures \ref{fig:ID1}a and \ref{fig:ID1}b, we have shown the resulting
patterns when  in the initial background pattern, the heights at a fraction $\epsilon$ of randomly
chosen sites are flipped from $1$ to $0$, and vice versa. The mean density of the background remains
$1/2$. The patterns correspond to $\epsilon = 0.01$ and $0.1$,
respectively. In this case, the most noticeable qualitative
changes seems to be the fact that boundaries between patches are no
longer sharp, which makes even a  precise definition  of a patch
difficult.

Comparatively, the patterns in an ASM on a undirected square lattice are
more robust against addition of small amount of randomness in background. We introduce randomness
in the uniform background of height $2$ by assigning each  site
heights $0,1$ or $2$ with probabilities $p/2,p/2$ and $1-p$, respectively,
independent of other sites. 

\begin{figure}
\begin{center}
\includegraphics[width=7.0cm]{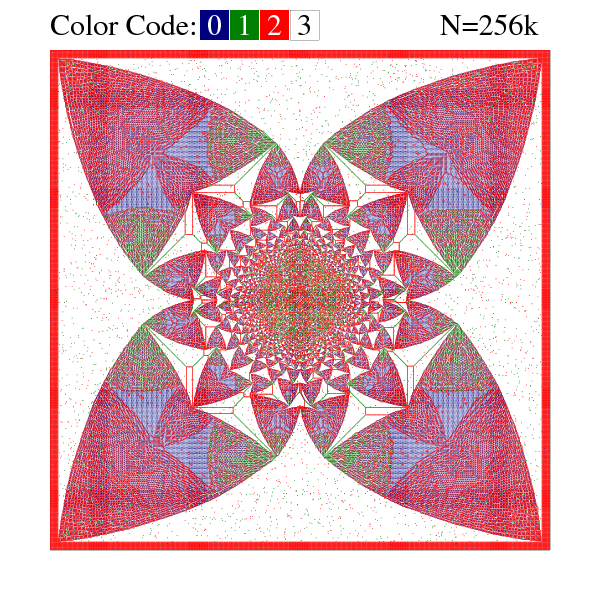}
\includegraphics[width=7.0cm]{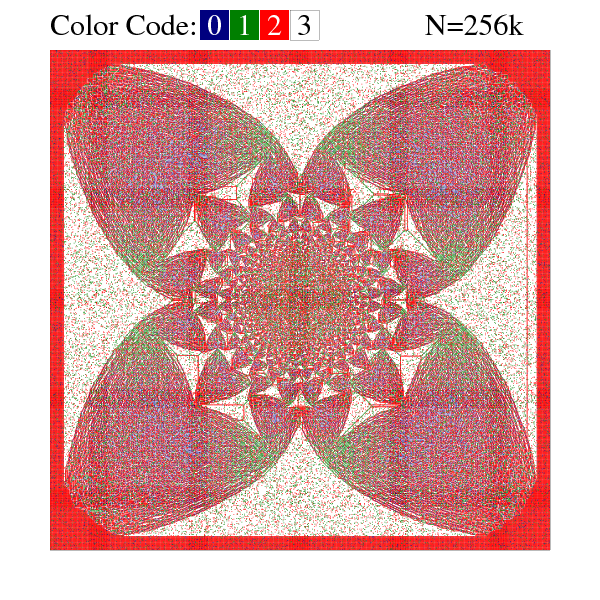}
\caption{The patterns produced on a square lattice with a background in
which heights $z=2$ at most of the sites except, $1\%$ and $10\%$ of the sites,
respectively, are with random assignment of heights $0$ or $1$. Color
code: Red=0, White=1.}
\label{fig:btwID}
\end{center}
\end{figure}

The patterns corresponding to $p=0.01$ and $p=0.1$ are shown in figure
\ref{fig:btwID}a and \ref{fig:btwID}b, respectively. These should be compared with the pattern produced
by adding $N=250,000$ grains on a background of height $2$ at all
sites, shown in figure \ref{fig:btw}.

At $p=0.01$, the patches with height predominantly
$3$, do not change much, except the presence of reduced heights at
the defect sites. This is easy to understand using an argument similar to the one
given for the F-lattice pattern. The presence of defect sites inside
the rest of the regions, generates line-discontinuities in the heights, which washes out
the smaller features of the pattern. As the noise is increased, the number
of defect lines increases, and the finer features are
lost. Also the corners of the outer boundary become round. At
noise strength $p\simeq 0.5$, the pattern becomes a circle with
random distribution of heights of uniform density, inside.

We note that, other types of the randomness in the initial conditions can have different 
behavior. For example, den Boer \textit{et.al.} \cite{denboer} have
shown that if one adds an arbitrary small density
of sites with height $3$, while all the other sites have height $2$ on the undirected square lattice,
one gets infinite avalanches for \textit{finite} $N$, with probability $1$. 

\section{Randomness in the Toppling matrix}\label{sec:quenched}

We now consider the effect of disorder in the underlying lattice on which the growth occurs.
We have considered two types of disorders. 
The first one is a bond disorder, where a randomly chosen fraction $\epsilon$ 
of the bonds are removed. Here, no grain transfer can occur along these bonds. For the undirected square lattice case, to keep the conservation law of sand in the model,
we change the critical height at the end points of each such  broken bond,
so that a site becomes unstable when its height equals
or exceeds its coordination number. The toppling rules are
deterministic, and the number of sand grains are conserved in a
toppling. However, the toppling matrix is no
longer translationally invariant.
\begin{figure}
\begin{center}
\includegraphics[width=7.5cm]{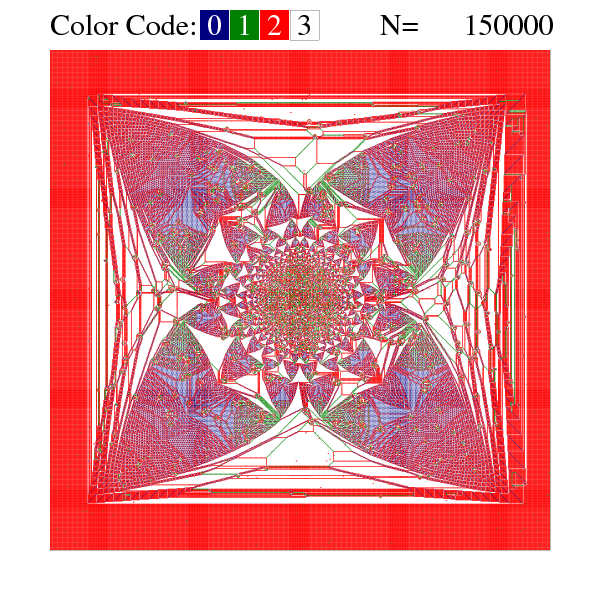}
\includegraphics[width=7.5cm]{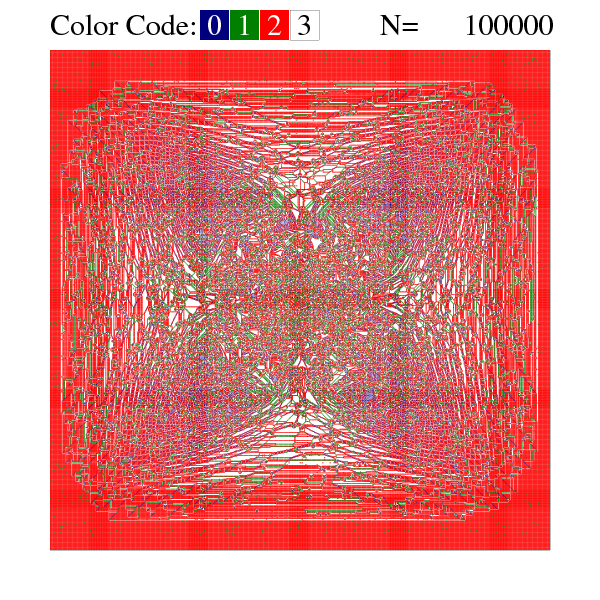}
\caption{The patterns produced on a square lattice with $1\%$ and
$10\%$ of the edges, respectively, are broken. Initial configuration with all heights $2$.}
\label{fig:btwBB}
\end{center}
\end{figure}

The patterns corresponding to the undirected ASM on the  square lattice with $1\%$ and
$10\%$ broken bonds are shown in figure \ref{fig:btwBB}. We see that
even for $\epsilon = 0.01$, the pattern has changed substantially. 
The outermost patches, which, in the absence of noise, had  all sites with height $3$, now show a large
number of criss-crossing defect lines. Further, counting inwards from
outside, one can clearly see at least
three or four more rings of patches. Fewer features are clearly
distinguishable, for larger noise.
However, some remnant of the
characteristic four-petal pattern of the noise-free case, can be
clearly seen even for $\epsilon=0.10$. 

\begin{figure}
\begin{center}
\includegraphics[width=8.0cm]{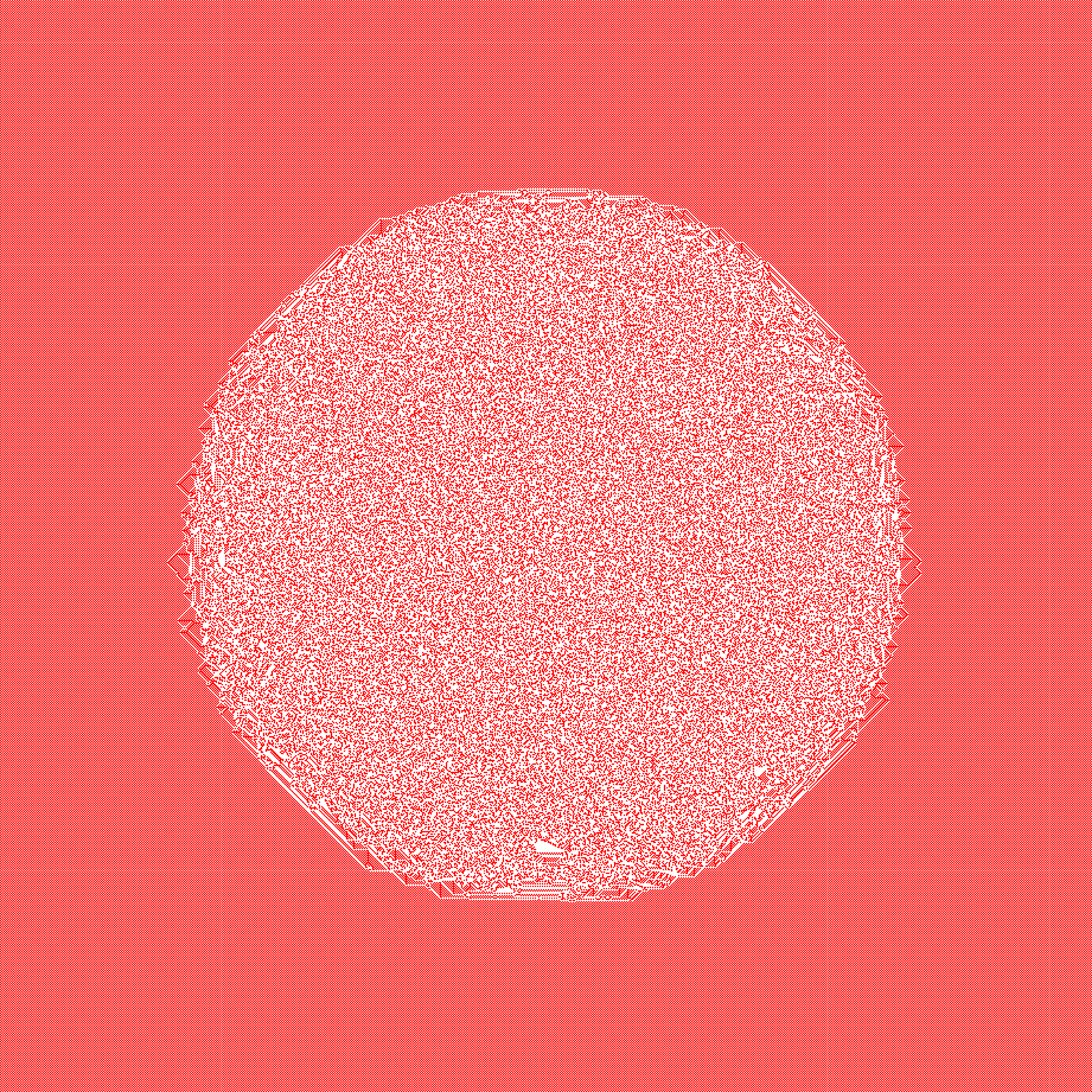}
\caption{A pattern produced on a F-lattice in which $1\%$ of the sites
has incoming and outgoing arrows switched. Color code: Red=0, White=1.}
\label{fig:rd}
\end{center}
\end{figure}

The second type of disorder that we considered, is a type of site-disorder. We consider the F-lattice, where 
a small fraction $\epsilon$ of the sites are chosen at random, and we change the direction of bonds 
going out (from up-down to left-right, and vice-versa). The critical
height remains unchanged, and is the same for all sites. However now,
at each site the number of in-arrows is not necessarily equal to the
number of out-arrows. As discussed by Karmakar \textit{et.al.}
\cite{karmakar}, this is a relevant perturbation,
and an arbitrarily small $\epsilon$ changes the critical exponents of
the
avalanches. We find that the patterns in growing sandpiles are also  unstable
to even a little amount of this kind of randomness. The pattern corresponding
to $p=0.01$ and $N =57,000$ is shown in figure \ref{fig:rd}. This pattern is a circle
with no distinguishable structures inside.
\section{Effect of randomness in the toppling}\label{sec:manna}
We have also studied the effect of noise in the toppling rules. We
have considered the F-lattice.
For each toppling at a site, the direction of the outgoing grains
differs from the direction of the outgoing arrows, with a probability
$\epsilon$.
The two grains go in the
direction of outgoing arrows with probability $1 -\epsilon$, while they go
in the direction of incoming arrows with probability $\epsilon$. 
The stochastic toppling rules take this modification of the ASM  to the 
Manna universality class, which is different from that of the
deterministic ASM with fixed toppling rules. In this case, we expect
that the pattern would  be unstable against such perturbations. This
is verified by our simulations.  We simulated the pattern
obtained by adding $57,000$ grains on the F-lattice with checkerboard
background and $\epsilon = 0.01$.  The resulting pattern
is a simple, nearly circular blob, with no other discernible features.
It is visually indistinguishable from the pattern in \fref{fig:rd}.

\section{Discussion}\label{sec:discussion}
The complicated and beautiful patterns produced in the growing
sandpiles are the result of an interplay between macroscopic conservation laws
(encoded in the Poisson's equation satisfied by the potential
function) and the integer nature of the microscopic variables. This
is not yet well understood. In fact, starting from the ASM rules, as yet we
can not prove even the existence of proportionate growth
in the growing patterns.

In the presence of noise, an analytical study of this problem is even
more difficult. Clearly, the potential function is no longer
piece-wise quadratic, and the analytical techniques used in
\cite{epl},
to characterize the pattern exactly, no longer work. In fact, in
figures \ref{fig:ID1} and \ref{fig:btwBB}, there are no sharp patch
boundaries, and perhaps one can not even give a clear definition of
`patches', at all. The patterns are characterized by the nontrivial
spatial dependence of the density function $\Delta \rho(\xi,\eta)$.
The pictures are reminiscent of Rayleigh-B\'{e}nard convection patterns
\cite{rayleigh}, where a linear analysis about the uniform steady state
shows that, in some regime of parameters, they
become unstable to a class of space-dependent perturbations.  In
our numerical studies, we see  that the featureless
circular-blob-pattern of growth at high noise levels, is unstable with
respect to
some characteristic low-wavelength density instabilities for low-noise
deterministic models. However, there is an important difference
between these two cases. In the convection problem, the amplitude of
the perturbations grows in time exponentially till it reaches a
saturation value, determined by the nonlinear terms, whereas in the sandpile
problem, the notion of ``growth of amplitude in time'' is not  well-defined.

Nevertheless, for the sandpile patterns there is the ``least action
principle'', which is a variational principle that allows us to compare
different trial patterns, and select the pattern corresponding to the
minimum action. Here `action' is measured in terms of the total number of 
toppling events. The
principle, in the ASM context, is informally stated as the lazy
man's maxim: ``Don't do anything, unless you have to''. If we think of
topplings as dissipative events, it is similar in spirit to the
principle of minimal heat production in resistor networks, or the
minimum entropy production principle, often discussed in
non-equilibrium statistical physics. While the extent of validity of
the latter, in general, is not clearly established (see, for example,
the discussion in  \cite{jaynes}), for this special case of ASM's
with a threshold rule for topplings, given that the order of
topplings does not matter, the principle is easily proved, and is more or
less built into the rules of evolution \cite{denboer}.

More precisely, if one considers a starting configuration  $C$ of an
ASM, with one or more unstable sites, then the toppling rules of the
ASM determine the stable final configuration $F$, uniquely. Suppose we
modify the toppling rules of the ASM by dropping the condition that a
toppling occurs only at sites where the height exceeds the threshold
value, and allowing topplings at any  site. For example, starting with a
configuration of all sites with height zero on  the undirected square
lattice, a toppling at the origin would make the height there $-4$,
and height at each of the neighbors, $+1$. Then, starting from $C$,
there is  a large number of  stable configurations reachable by
topplings. The `least action principle' for the ASM states
that, if we can reach a \textit{stable} configuration $F'$ from $C$
under this  less restrictive dynamics, the number of topplings
required to reach $F'$ is greater than that required to reach $F$, for
all $F' \neq F$. 

The variational principle allows us to compare  different guesses for
the final configuration, and tells us which one is closer to the actual
pattern. The main difficulty in applying this principle, in practice, is
that the set of configurations, over which the extremization has to be
done, is all possible configurations {\it reachable from the initial
unstable pattern by topplings}. Characterizing this set is rather
difficult. However, one can  restate this principle in terms of
non-negative integer toppling functions, $T_N(x,y)$. For any choice of
$T_{N}\left( x,y \right)$, there is a well-defined, easily computed, final height
configuration. Also, one can systematically improve on an initial
trial function, by performing more topplings at sites which are unstable in the final
configuration, or by untoppling at sites where the heights are too low.
This has been shown to yield a very efficient algorithm for
determining the final configuration for the related rotor-router
model \cite{tobias}.

There are other questions that have been addressed only partially in
this paper. Certainly, it would be useful to have a more quantitative
study of the patterns using the toppling function. Numerical studies
with significantly larger $N$, can clarify whether or not the density
function shows spatial discontinuities in the presence of low noise of the type shown in figures \ref{fig:idtype1} and
\ref{fig:btwID}, and whether the excess density is exactly a constant
within  a patch. It is hoped that further work will clarify these
issues.

We thank A. Libchaber for emphasizing the importance of introducing noise in our models, and Satya N. Majumdar for his
constructive comments on an earlier draft of this paper.
\section*{References}
\bibliographystyle{unsrt}    
   % (uses file "plain.bst")
\bibliography{mybib}
\end{document}